\documentclass[conference]{IEEEtran}
\IEEEoverridecommandlockouts
\usepackage{cite}
\usepackage{amsmath,amssymb,amsfonts}
\usepackage{algorithmic}
\usepackage{graphicx}
\usepackage{textcomp}
\usepackage{xcolor}
\def\BibTeX{{\rm B\kern-.05em{\sc i\kern-.025em b}\kern-.08em
    T\kern-.1667em\lower.7ex\hbox{E}\kern-.125emX}}
\begin{document}

\title{Towards training music taggers on synthetic data\\
}

\author{\IEEEauthorblockN{Nadine Kroher}
\IEEEauthorblockA{\textit{Time Machine Capital Squared Ltd.} \\
London, United Kingdom \\
nadine@tmc2.ai}
\and
\IEEEauthorblockN{Steven Manangu}
\IEEEauthorblockA{\textit{Time Machine Capital Squared Ltd.} \\
London, United Kingdom \\
steven@tmc2.ai}
\and
\IEEEauthorblockN{Aggelos Pikrakis}
\IEEEauthorblockA{\textit{Department of Informatics} \\
\textit{University of Piraeus}\\
Piraeus, Greece \\
pikrakis@unipi.gr}
\thanks{Aggelos Pikrakis is also a scientific advisor to Time Machine Capital Squared Ltd.}
}

\maketitle

\begin{abstract}
Most contemporary music tagging systems rely on large volumes of annotated data. As an alternative, we investigate the extent to which synthetically generated music excerpts can improve tagging systems when only small annotated collections are available. To this end, we release \textit{GTZAN-synth}, a synthetic dataset that follows the taxonomy of the well-known \textit{GTZAN} dataset while being ten times larger in data volume. We first observe that simply adding this synthetic dataset to the training split of \textit{GTZAN} does not result into performance improvements. We then proceed to investigating domain adaptation, transfer learning and fine-tuning strategies for the task at hand and draw the conclusion that the last two options yield an increase in accuracy. Overall, the proposed approach can be considered as a first guide in a promising field for future research.
\end{abstract}


\section{Introduction}
Systems which can automatically classify or tag music tracks are an essential component of large-scale music and multimedia indexing pipelines. Recent methods commonly rely on deep architectures which are either trained end-to-end on large annotated datasets (i.e. \cite{musicnn2}) or pre-trained on large volumes of real-world audio data and then fine-tuned for various downstream tasks on smaller collections (i.e. \cite{musicnn1}, \cite{maest}). Although both approaches have shown promising results, they still rely on large volumes of labelled data to effectively pre-train or train end-to-end deep neural networks, a prerequisite which comes at great cost. Manual labelling is a time-consuming and tedious task and crowd-sourced or user-contributed annotations are often noisy or inconsistent. To alleviate the problem, recent approaches commonly employ data augmentation strategies (see i.e. \cite{mcfee2015software}) which create variants of training instances by applying small modifications, i.e. pitch shifting or time stretching. While data augmentation does generally yield improvements, the augmented examples are still strongly correlated with the originals and do not prevent overfitting to the extent that adding new data would.

In an attempt to overcome the dependence on annotated data further, we explore an alternative strategy that makes use of synthetically generated music to train a tagging system. Similar approaches have been adopted by computer vision systems to address tasks \cite{man2022review} for which annotated data is hard to assemble, as it is for example the case with defect detection in the steel industry \cite{boikov2021synthetic}, vessel classification from overhead imagery \cite{ward2018ship} or medical image analysis \cite{kim2021deep}. In the audio domain, \cite{ronchini2024synthesizing} recently demonstrated that environmental sound classification can be improved by augmenting real-world datasets with synthetic examples generated with text-to-audio models.  

In order to explore a similar approach in the music domain, we recently conducted a preliminary proof-of-concept experiment which involved only a small taxonomy of five perceptually easy to distinguish genres \cite{kroher2023can}, which showed promising results. Based on these findings, we now extend our work to \textit{GTZAN} \cite{gtzan}, a small genre classification dataset, and explore various strategies for training on synthetic music data. Specifically, we first create \textit{GTZAN-synth}, a collection of artificially generated music excerpts that follows the \textit{GTZAN} taxonomy while being $10$ times larger in data volume. We make all code for reproducing and scaling this collection publicly available to foster further research into this direction. 

In addition, we explore several strategies towards improving the classification of a deep convolutional neural network on the \textit{GTZAN} dataset by incorporating synthetic music excerpts into the training procedure. Our experiments cover a wide spectrum of options, including the simple integration of the synthetically generated data with the training split of \textit{GTZAN}, the investigation of transfer learning and fine-tuning strategies, and a domain adaptation \cite{farahani2021brief} mechanism to mitigate the distributional shift between real and synthetic data. 
 
\section{Related work}
\subsection{Music tagging}
Music tagging systems analyse audio content in order to automatically annotate music tracks with  high-level descriptors related to mood (i.e. ``melancholic" or ``upbeat"), genre (i.e. ``rock" or ``jazz") or instrumentation (i.e. ``string quartet" or ``saxophone"). Most recent methods rely on deep neural network architectures, such as convolutional neural networks \cite{won2020evaluation} and transformers \cite{won2021semi}, which are usually trained end-to-end on large annotated music collections like the Million Song Dataset \cite{Bertin-Mahieux2011} or the Magna Tag-a-Tune dataset \cite{law2009evaluation}. Other approaches leverage representations learned in a supervised setting \cite{maest} on even larger datasets and then formulate specific tagging problems as downstream tasks which are solved via transfer learning. In both scenarios, large volumes of annotated data are required, which may not be available for every use-case.

In order to overcome this need, some recent efforts have focused on self-supervised techniques (partially \cite{li2022map} or fully \cite{castellon2021codified}) to learn audio representations. While this approach, which is also commonly used in natural language processing, has shown promising results, we explore in this paper the radically different idea of leveraging generative music systems to create artificial training data. To the best of our knowledge, with the exception of our previous exploratory study \cite{kroher2023can}, this is the first work aimed at investigating frameworks for training music taggers on synthetically generated music datasets.

\subsection{Generative music systems}
The recent advancements in generative artificial intelligence for natural language processing and image analysis tasks have also driven rapid improvements of state of the art systems that can generate music excerpts conditioned on text and melody prompts. Prominent generative music models include Jukebox \cite{dhariwal2020jukebox}, MusicLM \cite{agostinelli2023musiclm}, Jen-1 \cite{li2023jen} and MusicGen \cite{musicGen}. In this study, we focus on MusicGen, for which pre-trained models are publicly available via the \textit{Audiocraft}\footnote{https://github.com/facebookresearch/audiocraft/} library. 

While the distribution of synthetically created music via streaming services or its use in movies, TV or social media content poses legal issues \cite{smits2022generative} and has raised controversial discussions from an ethical point of view \cite{barnett2023ethical}, we believe that there is unexplored potential in leveraging such systems for the creation of synthetic training data for tagging and analysis models. 

\subsection{Domain adaptation}
Although generative systems for natural language and image content have become very sophisticated over the recent years and are now capable of producing convincing output, generative music systems, while advancing rapidly, are still limited in their ability to produce realistic music tracks and are highly relevant to the input prompt. Consequently, with a view on training music taggers, it is reasonable to expect a content distribution shift between real and synthetic data. 

This issue has also been addressed in other domains, mainly in computer vision, via the so-called domain adaptation (DA) methods, which essentially introduce additional loss terms during the training stage to ensure that a machine learning model trained partially or fully on synthetic data will generalise its performance to real world data. For a comprehensive review of different methods on various use-cases, training frameworks and domains, we refer to \cite{farahani2021brief}. In the context of music, existing work on domain adaptation has been limited to specific use-cases that do not involve synthetic data, i.e. cross-cultural emotion recognition \cite{chen2018cross} or the compensation of differences among microphones for the purposes of piano transcription \cite{bittner2022multi}. 

For the music tagging task at hand, we investigate the method proposed in \cite{motiian2017unified}, which addresses supervised DA in neural network training via an additional loss term that operates on a bottleneck layer and essentially acts as a contrastive loss that forces intermediate representations from real and synthetic data of the same class to be in close proximity in a Euclidean sense while maintaining large Euclidean distances to instances of different classes. 

\section{Data}
Our experiments make use of two datasets, the well-known \textit{GTZAN} \cite{gtzan} genre detection dataset and \textit{GTZAN-synth}, a synthetic dataset which we generated using the MusicGen \cite{musicGen} generative music system. \textit{GTZAN-synth} follows the \textit{GTZAN} class taxonomy but it is ten times larger. 

\subsection{Real music dataset}
During the early days of music tagging, the \textit{GTZAN} dataset was widely used as a benchmark collection to evaluate and compare competing approaches. It contains a total of $1000$ music excerpts of length $30s$ belonging to $10$ genres (e.g., ``blues", ``classical" or ``rock"), with each genre being represented by $100$ tracks. The dataset, including audio files and metadata, is publicly available\footnote{https://www.kaggle.com/datasets/andradaolteanu/gtzan-dataset-music-genre-classification} via the Kaggle platform.

In our experimental evaluation, we the use artist-filtered validation splits proposed in \cite{FOLEIS2020106127} which are available on GitHub\footnote{https://github.com/julianofoleiss/gtzan\_sturm\_filter\_3folds\_stratified/}.

\subsection{Synthetic music dataset}
In order to study the suitability of synthetically generated music for training tagging systems and foster future work on this topic, we release the synthetic music dataset \textit{GTZAN-synth}. Following the \textit{GTZAN} taxonomy, we create genre-specific prompts that are used to condition the medium-sized MusicGen model. Compared to the original \textit{GTZAN}, we scale by factor of $10$, thus generating $1000$ tracks per genre.

MusicGen is controllable via text prompts and while its generative process is to a certain degree stochastic, e.g. the prompt ``a rock song" will give different results if run repeatedly, we observed during our initial experiments that large volumes of excerpts generated with a single prompt do not exhibit sufficient diversity. Consequently, the main challenge in generating a synthetic music dataset lies in the creation of a large volume of genre-specific text prompts that yield sufficient variety while maintaining genre-specific characteristics. 

During initial experimentation, we also observed that MusicGen does not appear to generate vocals, even if prompted to do so. In addition, we noticed that prompts mentioning vocals often lead to output artefacts. As a result, we are restricted to generating instrumental examples. 

\begin{figure*}[h!]
\centerline{\includegraphics[width=\textwidth]{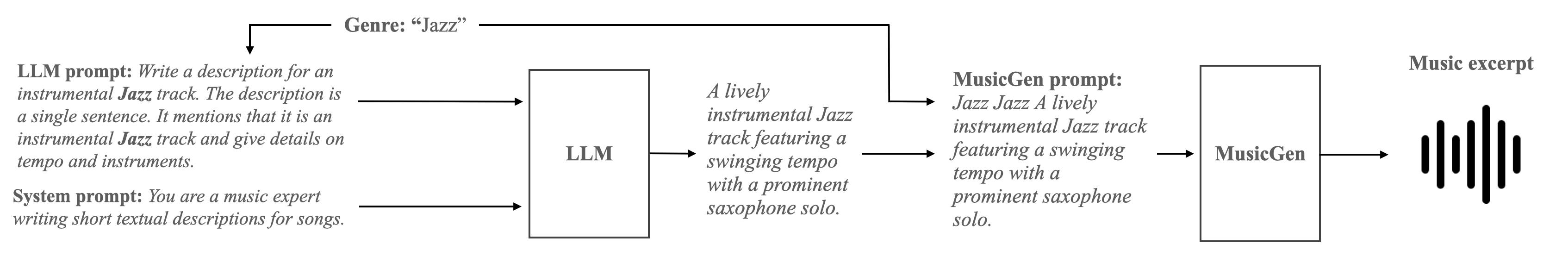}}
\caption{Schematic representation of the proposed prompt engineering pipeline. For a given music genre, we use an LLM to generate genre-specific track descriptions. After some additional processing, these serve as input to MusicGen, a model for text-conditioned music generation.}
\label{fig2}
\end{figure*}

To assemble an adequate number of text prompts for the MusicGen synthesis engine, we used a large language model (LLM) of the GPT-3 \cite{floridi2020gpt} family which can be accessed via the OpenAI API\footnote{https://openai.com/blog/openai-api}. To avoid confusion, we will refer to the text input that drives the GPT-3 LLM as the \textit{LLM-prompt} and the text guidance input to the MusicGen model as the \textit{MusicGen-prompt}. Figure \ref{fig2} depicts our prompt engineering pipeline along with an example. 

We make the recipe for generating \textit{LLM-promts}, the generated $10$k \textit{LLM-promts}, as well as code for generating the music excerpts with MusicGen publicly available\footnote{URL will be inserted upon paper acceptance}.

\subsubsection{LLM-prompt engineering}
We used the \textit{gpt-3.5-turbo} model available via OpenAI's chat completion API and set the temperature parameter to $0.5$ to ensure sufficient variety over consecutive runs. The model takes an optional system prompt as an additional input which steers the overall behaviour of the model with respect to tone and style of interaction. Here, we used the following short system prompt: \textit{You are a music expert writing short textual descriptions for songs.} When engineering the \textit{LLM-prompt}, we aimed at enforcing the generation of prompts that are limited to instrumental music, that mention the genre specifically and that describe various aspects of a music track. We experimentally found that the following \textit{LLM-prompt}, where \{genre\} is a placeholder for the respective genre, yielded stable and convincing results: \textit{Write a description for an instrumental \{genre\} track. The description is a single sentence. It mentions that it is an instrumental \{genre\} track and gives details on tempo and instruments.} 

\subsubsection{MusicGen-prompt engineering}
While most of the generated textual descriptions yielded convincing output when used to guide MusicGen, we observed that simply mentioning the genre once in the prompt did not guarantee a good alignment with the target class. We experimented with different prompt engineering strategies and observed that the best option for forcing the model to generate music excerpts that are prototypical for a particular genre is to simply prepend the genre several times to the prompt. 

Based on this observation, we generated the \textit{MusicGen-prompts} by concatenating two repetitions of the class name with the \textit{LLM-promp} as follows: \textit{\{genre\} \{genre\} \{LLM-prompt\}}. 

\subsubsection{Music generation}
Finally, in order to generate synthetic music excerpts, we employ the medium-sized version of the pre-trained MusicGen model, which has around $1.5B$ parameters, and condition it on the \textit{MusicGen-prompts}. The model operates on top of the EnCodec tokenizer \cite{defossez2022high} which decodes tokens from $4$ codebooks at a sample rate of $50$ Hz to raw audio with a sampling rate of $32$ kHz.

\section{Model Architecture}
Since the aim of this paper is not to propose a novel genre tagging method but rather to explore the potential of using synthetically generated data, our neural network design roughly follows the MusiCNN architecture described in \cite{musicnn2} which was intended for end-to-end training of music taggers on large volumes of data.  

\begin{figure*}[h!]
\centerline{\includegraphics[width=\textwidth]{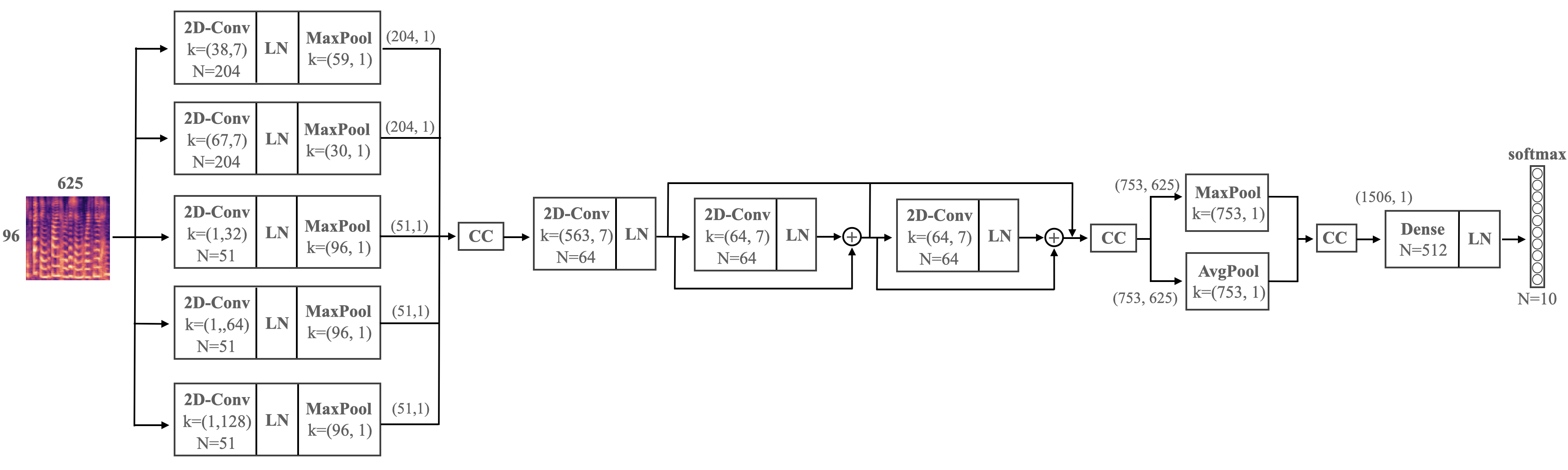}}
\caption{Schematic representation of the model architecture which takes a mel-spectrogram as input and predicts the music genre in a multi-class classification task. \textbf{2D-Conv}: two-dimensional convolutional layer. \textbf{LN}: layer normalization. \textbf{CC}: concatenation. \textbf{MaxPool}: maximum pooling. \textbf{AvgPool}: average pooling.}
\label{fig3}
\end{figure*}

The network operates on $96$-band mel-spectrograms extracted from $10$s of audio with a sampling rate of $16$kHz, a window size of $512$ samples and a hop size of $256$ samples. The input is then processed by a first layer of parallel convolutional filters of different kernel shapes which are designed to learn different temporal and timbral features. The resulting feature maps are then pooled across the full frequency axis, concatenated and processed by further convolutional layers with residual connections and pooling operations before being passed through a normalised dense layer with $512$ units and a final softmax classification layer with $10$ units corresponding to the $10$ genre classes. The architecture is shown in Figure \ref{fig3}. For a detailed description of the motivation behind the design choice the reader is referred to \cite{musicnn2}.

In our experimental evaluation, we train the model end-to-end and also in transfer learning and fine-tuning settings. In the case of transfer learning, the entire network is pre-trained on a large dataset and then only the penultimate dense layer and the final classification layer are trained from scratch on the smaller dataset with all other layers keeping their weights frozen. In the fine-tuning scenario, the entire network is first trained on a large collection and then training for all weights is resumed on the smaller collection. 

\section{Supervised domain adaptation}
\label{sec-da}
In order to compensate for potential distributional discrepancies between synthetic and real music data, we experiment with the supervised domain adaptation approach proposed in \cite{motiian2017unified}. We can describe the genre classification network as $f = g \circ h$, where $g: \mathcal{X} \rightarrow \mathcal{Z}$ takes the mel-spectrogram $\mathcal{X}$ as input and maps it to the penultimate dense layer $\mathcal{Z}$, which we will treat as an intermediate feature representation. The classification layer $h: \mathcal{Z} \rightarrow \mathcal{Y}$ maps this intermediate representation $\mathcal{Z}$ to the final classifier output $\mathcal{Y}$. 

Without any counter measures, any distributional shift between real $X_r$ and synthetic $X_s$ inputs can potentially propagate to the intermediate representation, resulting in discrepancies between $\mathcal{Z}_r$ and $\mathcal{Z}_s$. In order to drive the network towards learning feature representations which encode commonalities between the two domains, we employ a contrastive loss $\mathcal{L}_{SA}$ in addition to the classification loss $\mathcal{L}_{CLS}$. $\mathcal{L}_{SA}$ encourages semantic alignment rather than domain-specific alignment of the intermediate feature representation. 

Given a synthetic instance $x_s^i$, we can compute the contrastive loss by comparing its embedding to that of a real sample $x_r^j$ of the same label and to that of a real sample $x_r^k$ of a different label, where $y^i = y^j$ and $y^i \neq y^k$ as follows:

\begin{equation}
    d(g(x_s^i, x_r^j) = \frac{1}{2} || g(x_s^i) - g(x_r^j) ||^2
\end{equation}

and 

\begin{equation}
    d(g(x_s^i, x_r^k) = \frac{1}{2} \max(0, m - || g(x_s^i) - g(x_r^j) ||^2)
\end{equation}

where $m$ denotes a margin parameter related to the separability in the embedding space and the semantic alignment loss becomes $\mathcal{L}_{SA} = d(g(x_s^i, x_r^j) + d(g(x_s^i, x_r^k)$. In practice, we did not observe a difference between comparing to a single randomly selected target instance versus computing the average distance over all target samples as originally suggested by \cite{motiian2017unified}. The second option is computationally very inefficient since the forward pass over all target samples has to be computed at each training step. 

Finally, both loss terms can be combined as $\mathcal{L} = \gamma \mathcal{L}_{SA} + (1- \gamma) \mathcal{L}_{CLA}$, where $\gamma$ is a balancing factor which controls the relative weight of each term.

\section{Experimental evaluation}
In order to assess the potential of using synthetically generated music for training tagging systems, we run a series of experiments. For all settings, we train with a batch size of $4$ and use the Adam \cite{kingma2014adam} optimizer with an initial learning rate of $0.001$, except for the fine-tuning setup where we decrease the initial learning rate to $0.0001$. For the domain adaptation experiment, we set the margin to $m=2$ and the balance parameter to $\gamma = 0.7$. We furthermore use an early stopping criterion based on the classifier's categorical cross-entropy loss with a patience period of $5$ epochs and select the best performing model for evaluation. For the GTZAN dataset, where excerpts are of length $30$s, we use a random $10$s excerpt during training and the central $10$s segment during validation.

For all experiments evaluated on the GTZAN dataset, we report mean and standard deviation of the categorical accuracy and of the categorical cross-entropy loss over the three validation splits proposed in \cite{FOLEIS2020106127}. When evaluating on synthetic data, we report mean and standard deviation over three random splits instead. All results are shown in Table \ref{tab1}

\begin{table}[htbp]
\caption{Results of the experimental evaluation. We report mean $\mu$ and standard deviation $\sigma$ of the categorical accuracy and of the categorical cross-entropy loss over the three artist-filtered validation splits proposed in \cite{FOLEIS2020106127}.}
\begin{center}
\begin{tabular*}{\linewidth}{@{\extracolsep{\fill}} cccc }
\hline
\textbf{System} & \textbf{Accuracy} $\mu (\sigma)$ & \textbf{Loss} $\mu (\sigma)$\\
\hline
E2E-real & $46.7\% (5.2\%)$ &$1.61 (0.05)$ \\
E2E-synth & $90.6.\% (1.6\%)$ &$0.34 (0.02)$ \\
E2E-add & $47.3\% (4.6\%)$ &$1.74 (0.05)$ \\
E2E-DA & $47.6\% (6.0\%)$ &$1.54 (0.06)$ \\
TL & $52.6\% (3.3\%)$ &$1.40 (0.06)$ \\
FT &$54.8\% (6.2\%)$ &$1.39 (0.12)$ \\
\hline
\end{tabular*}
\label{tab1}
\end{center}
\end{table}

First, we train and evaluate the model solely on the \textit{GTZAN} dataset (\textbf{E2E-real}). Given the relatively large number of network parameters and the rather small dataset volume, we observe that the model overfits early and yields a mean accuracy of $46.7\%$. We also train and evaluate the same model on the larger \textit{GTZAN-synth} dataset (\textbf{E2E-synth}) and observe a much higher mean accuracy of $90.6\%$. 

While the result on the synthetic data appears encouraging, it does not guarantee that this classifier will generalise well to real-world music excerpts. It is for example possible that the synthetic music excerpts only represent a narrow fraction of the distribution of real music examples of the same genre. Another possibility is that the generative model causes artefacts or characteristics that are not found in real-world examples, thus causing distributional discrepancies between real and synthetic data. The fact that such issues do indeed exist is confirmed when we repeat the first experiment and simply add \textit{GTZAN-synth} to the training splits of \textit{GTZAN} (\textbf{E2E-add}). Here, we observe that the mean accuracy on the \textit{GTZAN} validation does not seem to improve ($47.3\%$) and the loss even slightly worsens ($1.74$ vs. $1.61$). 

\begin{figure}[htbp]
\centerline{\includegraphics[width=0.4\textwidth]{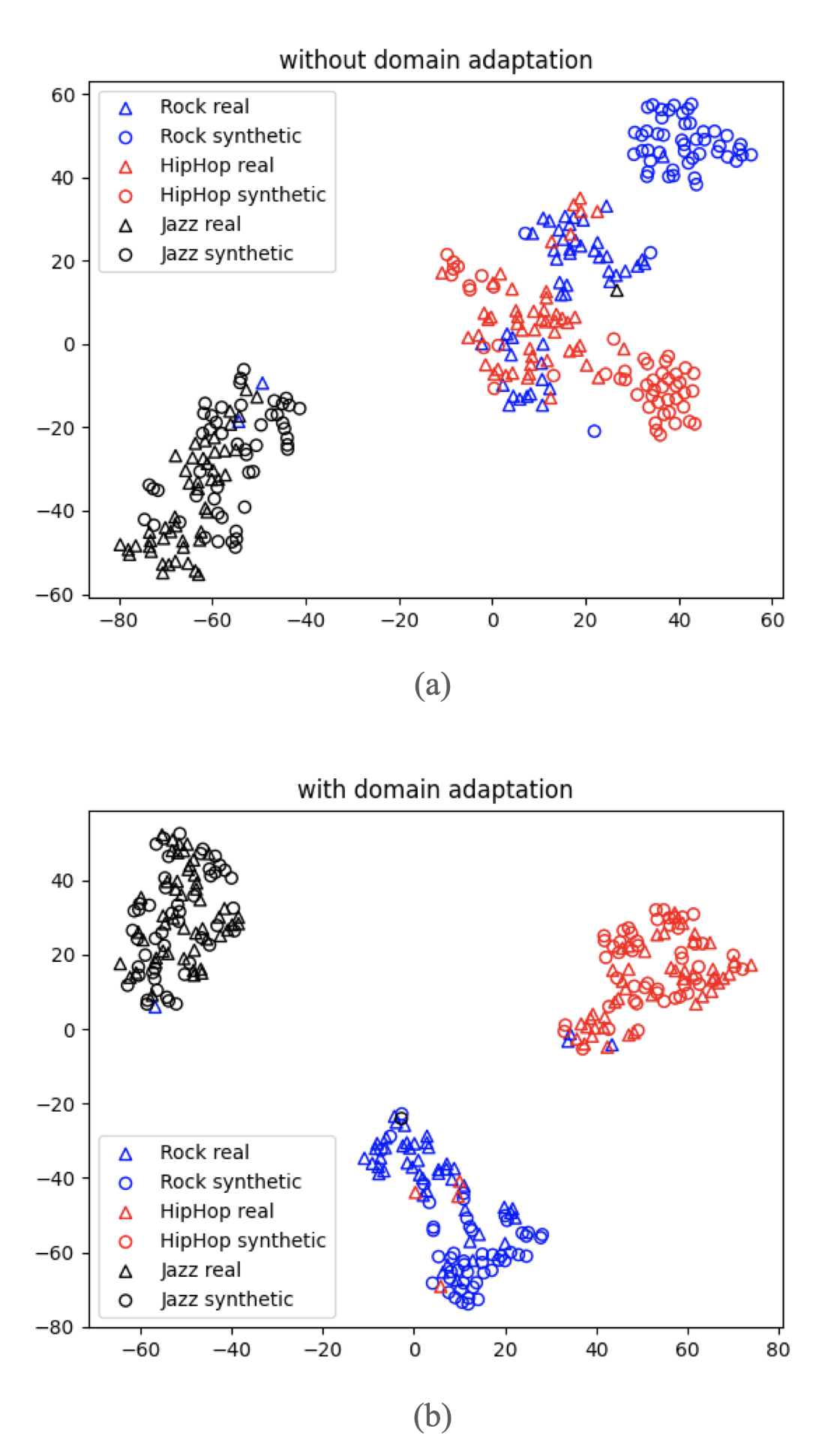}}
\caption{t-sne visualisations of intermediate representations of a subset of real and synthetic data with (bottom) and without (top) DA for three genres.}
\label{fig4}
\end{figure}

In an attempt to mitigate any distributional discrepancies between real and synthetic data, we repeat the previous experiments but employ the DA method (\textbf{E2E-DA}) described in Section \ref{sec-da}. We first visualise the intermediate representations after the penultimate layer with and without the additional DA loss term by approximating a two-dimensional representation using the t-sne \cite{van2008visualizing} algorithm. The resulting plot in Figure \ref{fig4} (a)) reveals that there does indeed appear to be a distributional discrepancy between real and synthetic data which seems to largely disappear when DA is used (Figure\ref{fig4} (b)). However, the results do not indicate a significant improvement in classification accuracy ($47.6\%$) and only the loss appears to have somewhat improved ($1.54$). One reason for this behaviour could be that by driving synthetic data towards the real-world examples, the lack of variety in the training dataset persists and thus generalisation does not improve.  

As a next step, we investigate if the model trained solely on synthetic data (\textbf{E2E-synth}) can be adapted for to real world data via transfer learning or fine-tuning. In the transfer learning experiment (\textbf{TL}), we freeze the convolutional layers from \textbf{E2E-synth} and only train the last two layers on \textit{GTZAN}. For this setup, we observe an increase in mean accuracy to $52.6\%$. When we fine-tune on \textit{GTZAN} by initialising the network with the weights from \textbf{E2E-synth} and by resuming training on all layers, we obtain an even slightly higher increase in mean accuracy to $54.8\%$. This indicates that training on synthetic data allows the network to learn useful features that aid in learning downstream tasks.

While these last two experiments demonstrate that synthetic data can help to improve performance of deep architectures on small datasets, it is worth noting that the results still do not reach those reported for systems that rely on transfer learning from large annotated datasets (i.e. $82.1\%$ reported by \cite{lee2018samplecnn}) or more lightweight end-to-end systems (i.e. $ 65.8\%$ reported by \cite{medhat2017masked}).

\section{Conclusions}
We have investigated the use of artificially created music excerpts for training tagging systems on the example of a genre classification task. To this end, we release \textit{GTZAN-synth}, a collection of synthetic music excerpts that follows the taxonomy of the well-known \textit{GTZAN} dataset but is $10$ times larger. In a series of experiments, we first demonstrated that simply adding synthetic data during training does not yield a significant performance improvement. We furthermore investigated domain adaptation, transfer learning and fine-tuning and showed that the two last options are suitable methods for increasing the performance of deep architecture trained on small datasets. We believe that these findings together with the release of \textit{GTZAN-synth} can foster future research into this direction, for example on advanced prompt engineering for generative music systems, on the influence of scaling the amount of synthetic data or on extending this work to other custom tagging tasks.


\bibliographystyle{IEEEtran}
\bibliography{conference_101719}


\end{document}